\documentclass[sigconf]{acmart}
\usepackage{longtable}

\AtBeginDocument{%
  }

\setcopyright{acmlicensed}
\copyrightyear{2025}
\acmYear{2025}
\acmDOI{10.1145/3698061.3726923}
\acmConference[C\&C '25]{Creativity and Cognition}{June 23--25,
  2025}{Virtual}
\acmISBN{979-8-4007-1289-0/25/06}

\begin{document}

\title[Crafting a Personal Journaling Practice]{Crafting a Personal Journaling Practice: Negotiating Ecosystems of Materials, Personal Context, and Community in Analog Journaling}

\author{Katherine Lin}
\affiliation{%
  \institution{Columbia University}
  \city{New York}
 \state{New York}
   \country{USA}
 }
 
\email{kl3085@columbia.edu}
\authornote{Both authors contributed equally to the paper}

\author{Juna Kawai-Yue}
\affiliation{%
  \institution{Columbia University}
  \city{New York}
 \state{New York}
   \country{USA}
 }
\email{jk4141@columbia.edu}
\authornotemark[1]

\author{Adira Sklar}
\affiliation{%
 \institution{Barnard College}
 \city{New York}
\state{New York}
  \country{USA}
}
 \email{as7172@barnard.edu}

 \author{Lucy Hecht}
\affiliation{%
 \institution{Columbia University}
 \city{New York}
\state{New York}
  \country{USA}
}
 \email{lbh2148@columbia.edu}

  \author{Sarah Sterman}
\affiliation{%
 \institution{University of Illinois Urbana-Champaign}
 \city{Urbana}
\state{Illinois}
  \country{USA}
}
 \email{ssterman@illinois.edu}

   \author{Tiffany Tseng}
\affiliation{%
 \institution{Barnard College}
 \city{New York}
\state{New York}
  \country{USA}
}
 \email{ttseng@barnard.edu}

\begin{abstract}

Analog journaling has grown in popularity, with journaling on paper encompassing a range of motivations, styles, and practices including planning, habit-tracking, and reflecting. Journalers develop strong personal preferences around the tools they use, the ideas they capture, and the layout in which they represent their ideas and memories. Understanding how analog journaling practices are individually shaped and crafted over time is critical to supporting the varied benefits associated with journaling, including improved mental health and positive support for identity development. To understand this development, we qualitatively analyzed publicly-shared journaling content from YouTube and Instagram and interviewed 11 journalers. We report on our identification of the \textbf{journaling ecosystem} in which journaling practices are shaped by \textbf{materials}, \textbf{personal context}, and \textbf{communities}, sharing how this ecosystem plays a role in the practices and identities of journalers as they customize their journaling routine to best suit their personal goals. Using these insights, we discuss design opportunities for how future tools can better align with and reflect the rich affordances and practices of journaling on paper. \end{abstract}



\maketitle
\section{Introduction}
\label{sec:intro}

Putting thoughts to the page is a storied practice, with physical notebooks playing a critical role in supporting idea-construction \cite{abtahi2020understanding,inie2022researchers}, memory \cite{cardell2017modern}, and historical preservation for centuries \cite{allen2024notebook}. The practice of journaling in particular, or the autobiographical recording of one's experiences, thoughts, aspirations, and ideas, has a long history of supporting well-being and self-reflection \cite{ayobi2018flexible, baldwin1977one, michael2003journal, rainer1979new, yuan2021think}. Despite a wide array of digital tools available today for note-taking and journaling, there has been a recent revitalization of analog journaling on paper, with the rise of Japanese stationery products in western markets \cite{japantimes} and heavily accessorized decorative planning \cite{wired}, especially among Gen Z and millennials \cite{guardian}. Hashtags like \#journaling and \#junkjournal, along with well-beloved notebook brands like \#hobonichi and \#travelersnotebook, each have over 1 million posts on Instagram, indicating the popularity and strength of active users engaging with journaling today. 

Recent efforts studying journaling practices in HCI have focused on data-driven practices, such as users who engage in lifelogging in the `quantified-self' movement \cite{abtahi2020understanding, ayobi2018flexible}, as well as use of digital notetaking applications \cite{elsden2016s, grahame2016digital,shi2020effects,katayama2005promoting} and blogging \cite{davis2010coming,boniel2013therapeutic,celdran2019older}. These efforts have examined user motivations for journaling and how well these motivations are supported by both analog and digital practices. In contrast, our research is primarily concerned with how user's journaling practices \textit{develop over time} and the role of factors beyond the page in shaping journaling practice. Our goals are to, 1) characterize the attributes and factors that contribute to how an individual crafts their own journaling practice, 2) identify design considerations for how these factors might be better accounted for in future journaling tools.

To uncover user practices and preferences when engaging with analog journaling over time, we utilized data across three sources: YouTube videos of journalers sharing their favorite stationery tools; Instagram posts sharing a diversity of journaling layouts; and interviews with active journalers in which they shared their physical journals and described how their journaling routine and goals develop. These data sources enable us to consider how material choices, such as what writing instruments and accessories are used, manifest on the page and relate to an individual's decisions about what their journal represents about their life.

Through this analysis, we situated analog journaling in a larger ecosystem of \textit{materials}, \textit{personal context}, and \textit{community} that each contribute to how journalers choose how to journal and what to journal about. We then discuss design opportunities for how changing practices over time can be more fully considered in the design of future tools to support journalers. Overall, our contributions are:
\begin{enumerate}
    \item A framework of a journaling ecosystem and description of how each component of this ecosystem affects journaling practices
    \item Characterization of how journaling practices are crafted and formed over time and identifications of shortcomings and opportunities for journaling tools and technologies to better align with these practices
\end{enumerate}

\section{Related Work}
\label{sec:related-work}

Our research draws from literature in the space of personal documentation more broadly to inform our understanding and interpretations of modern journaling practices, given how central documentation is to both the act of writing and motivations for doing so. This related work section covers two areas: 1) an overview of documentation practices and preferences across domains, and 2) tools that blend digital and analog approaches to supporting documentation.

\subsection{Documentation Practices and Preferences}

Personal documentation practices span diverse domains and user motivations \cite{allen2024notebook}. For example, documentation and various forms of journaling are a popular practice for supporting physical health \cite{luo2021foodscrap,cordeiro2015barriers,anhoj2004feasibility}, mental well-being \cite{smyth2018online,jean2024creative,langan2022reflective}, and mindfulness \cite{ayobi2018flexible, ramasubramanian2017mindfulness}. Personal documentation plays a critical role for creative practices, including artistic expression \cite{li2021we, sterman2022towards} and idea management \cite{inie2022researchers,rosselli2024my,dalsgaard2023capturing}. It is commonly associated with user goals of capturing and preserving one's legacy \cite{gulotta2013digital} and daily life through life-logging \cite{abtahi2020understanding, choe2014understanding,elsden2016quantified,elsden2016s}. 

In the context of analog documentation, prior research sheds light on the benefits of reflective writing on paper, which offers an open format that can foster self-discovery \cite{yuan2021think, davis2010coming}. For instance, studies on self-tracking and bullet journaling, which incorporate habit, health, and wellness tracking, show that users value the slow process provided by paper, which enables reflection, relaxation, and the ability to reminisce \cite{abtahi2020understanding}. Additionally, people appreciate the flexibility of paper \cite{luo2021foodscrap, ayobi2018flexible} and its finite space, which encourages prioritization \cite{tholander2020crafting, abtahi2020understanding, fernandez2020live}. Further, prior work has shown how people value the natural wear and tear of materials as they convey evolution and context, such as homemade cookbooks \cite{davis2014homemade} and heritage artifacts \cite{gulotta2013digital}, suggesting that meaning lies not only in content but also in its physical form. 

People who document on paper frequently incorporate digital tools into their practices \cite{steimle2011integrating,inie2022researchers}, and the moments when they opt for digital over analog underscore the distinct roles each medium fulfills \cite{dalsgaard2023capturing}. Shouzhang et al. distinguish between ``slow media,'' referring to reflective analog journaling, and complementary ``fast media,'' referring to social media, which enables sharing and community engagement \cite{yuan2021think}. A similar model appears in researchers’ documentation, as studies show that researchers often transition away from analog tools once they move beyond key ideas, shifting toward more complex, hyperlinked structures that support collaboration\cite{inie2022researchers}.

Individuals who partake in self-tracking often collect health data digitally before copying it down in analog form \cite{tholander2020crafting, abtahi2020understanding}. Similarly, artists, despite valuing creative freedom, also seek efficiency and the benefits of digital tools like copy-paste, undo and redo, and version control, features that facilitate managing and revisiting past drafts and foster exploration\cite{sterman2022towards, li2021we}. These preferences for flexible and lightweight tools are reflected in systems like list.it, which supports the quick capture of informal notes—tasks, reminders, links—without requiring structure or categorization \cite{vankleek2009notetoself}.

Our work builds on this broader understanding of documenting practices and preferences, which highlights the unique benefits of both analog and digital tools. By examining how individuals engage with their journaling over time, we aim to contribute to how future journaling tools might sustain positive journaling routines and practices. 

\subsection{Bridging Analog Documentation with Digital Affordances}
Researchers and designers have developed a variety of hybrid tools that integrate analog documentation with digital affordances. These tools aim to enhance efficiency and workflows \cite{innocent2024optimizing,abtahi2020understanding}, collaboration \cite{hill1992edit, bodker2024material}, and creativity \cite{ceh2024creativity, sterman2022towards} without sacrificing the tactile and reflective aspects that make physical practices meaningful.

Edit Wear and Read Wear technologies, for example, mimic the ``wear and tear'' of physical documents by tracking online interactions and visually mapping engagement over time \cite{hill1992edit}. Similarly, OneTrace, a cross-application history tool, unifies different collaborators’ historical data to help them track and coordinate past interactions \cite{battut2024onetrace}. Designed for collaboration, these text editors highlight changes to reveal revision patterns and authorship, allowing for reflection and explorative iteration. Alongside collaborative tools, systems like Scraps reflects a growing interest in tools that seamlessly connect fragmented, mobile capture with more structured reflection and documentation processes, enabling users to capture diverse digital content on mobile devices and later re-integrate it into text editors \cite{swearngin2021scraps}. 

Just as digital tools incorporate analog qualities, analog tools can be enhanced with digital capabilities. DigitalDesk projects a computer display onto a physical surface and uses cameras to track user interactions, granting paper documents digital features such as scanning handwritten numbers and instantly running calculations \cite{wellner1993interacting}. ButterflyNet similarly augments paper notebooks with digital capture tools like smart pens and cameras, enabling field biologists to document notes, photos, and specimens without abandoning analog methods \cite{yeh2006butterflynet}. These integrations show a path towards bypassing the manual transfer of digitally tracked data into analog materials, preserving familiar interactions while increasing efficiency and elasticity.

Increasing with the rise of generative AI, researchers are exploring how AI can support journaling practices. Customized prompts and chat-based diaries are a frequent approach; from using contextual data to generate diary prompts \cite{gyeyoung2024mylistener}, to conversational interactions with personal journaling \cite{schulz2024impact}, mental health support \cite{kim2024mindfuldiary} and workplace logging and reflection \cite{Kocielnik2018designing}, these tools seek to add interactivity and personalization to the journaling experience.
Like paper journals with prompts and guides, there is a relatively recent rise of the genre of prompt-based ``self-help journaling guides'' \cite{whitney2005writing}.
Yet one major concern with this approach is that journaling with an AI may unintentionally shape the writers' own thoughts and feelings, changing how they process their emotions \cite{kim2024diarymate}. 

 For many, journaling is not just about creating a final product but about the act of writing itself--getting thoughts down on paper and processing emotions and tasks \cite{ayobi2018flexible, abtahi2020understanding}. Apps like Spyn embrace this process-oriented approach by documenting the crafting journey in addition to the final product \cite{rosner2010spyn}. Spyn embeds digital traces into needlecraft, allowing crafters to pin thoughts and updates throughout their work and encouraging others to value their progress and the entire creative process. Similarly, our research is concerned with the development of practices over time and how they might be better supported in the domain of journaling. Through our formative research on understanding modern journaling routines that integrate both digital and physical affordances, we offer several pathways for future integration.  %
\section{Methodology}
\label{sec:methodology}

To explore contemporary practices for journaling (including user preferences for analog journaling tools, practices for laying out spreads (layouts across two pages), and developing journaling routines) we combined content analysis of publicly shared journaling videos and posts on social media with semi-structured interviews with journalers. 

For our analysis of public journaling content, we selected YouTube videos and Instagram posts as our data sources due to their popularity within journaling communities. Our analysis of YouTube content primarily focused on \textit{tool selection}, as walk through videos of what writing instruments and journals people use are a popular format. We used Instagram posts to study diverse examples of journaling spreads. Finally, we used our interviews to provide in-depth context for motivations and practices for journaling that were often not addressed directly in social media posts. In this section, we describe our content analysis approach for analyzing each of these distinct threads of data, along with how we combined our analyses to inform our insights on journaling practices.

\subsection{YouTube Videos on Journaling Tools}
\textbf{Data Collection}. To examine how journalers choose their writing tools, we collected and analyzed YouTube videos on journaling tool selection, including walkthroughs in which journalers describe their writing tools, decorative stationery, and journals. YouTube videos were selected using search queries that include, ``What's in my pencil case'', ``What's in my pouch'', and ``What's in my everyday carry,'' which are popular titles of journaling videos describing writing tools and accessories (`everyday carry' or EDC videos generally describe tools consistently carried and used by an individual).

We selected full YouTube videos (as opposed to shorter YouTube Shorts) based on diversity of content and metadata such as the videos' view counts and the number of subscribers for each channel. We selected a total of 29 YouTube videos published by 26 unique accounts, providing a total of about 6 hours of video footage with an average duration of around 13 minutes. The videos had a mean view count of 156k views each (SD = 335.3k). %
For full details of the selected videos, please refer to Table \ref{tab:youtube-videos}. 

\begin{figure*}[htb]
    \begin{center}
        \includegraphics[width=\textwidth]{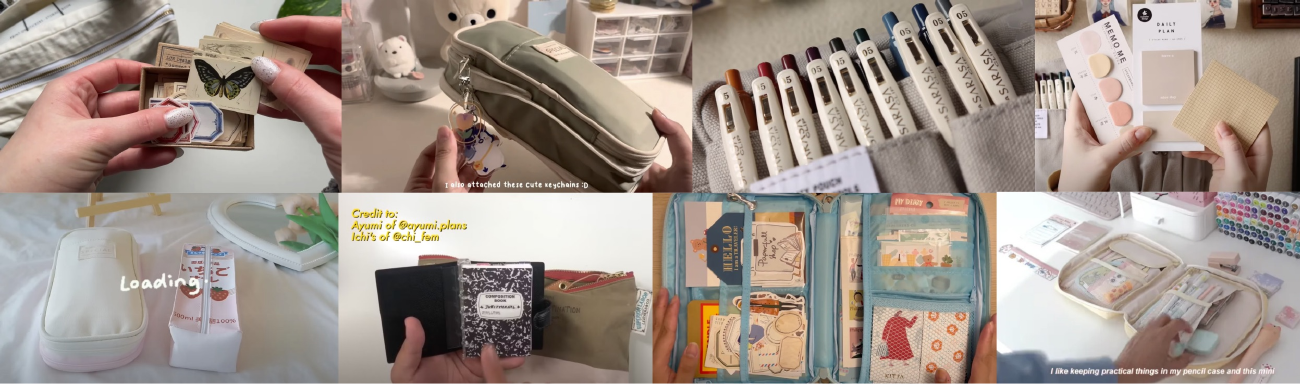}
        \captionof{figure}{Screenshots of YouTube videos analyzed for writing tools and accessories (clockwise from top left: $@$hellysjournal, $@$leighpadme, $@$KaitlinGrey, $@$KaitlinGrey, $@$miicchaa, $@$onjerraslist, $@$JobsJournal, $@$ErinFlotoDesigns)}
        \Description{Eight screenshots of YouTube videos displaying journaling tools} %
        \label{fig:youtube-spreads}
    \end{center}
\end{figure*}

\textbf{Analysis}. Our approach for analyzing YouTube videos (as well as the Instagram and interview data) was to utilize thematic analysis \cite{braunUsing2006} to qualitatively analyze the data and collectively identify core themes around user choice and preferences. To begin, three researchers independently watched the same 6 YouTube videos and applied open coding to both the transcript and visual content, taking note of depicted writing tools and the creators' explanations of why those tools were preferred. After discussing and agreeing upon common codes as a group, the researchers used affinity diagramming (using Miro) to begin categorizing and organizing the codes, using a combination of quotes from the videos along with stills showing visual characteristics of selected tools. The remaining YouTube videos were then split among 2 researchers, with each researcher watching around 10 videos independently to add and refine our affinity diagram. Our affinity diagram was discussed and constructed in weekly meetings over a duration of 3 months, and consolidated into a codebook of identified preferences and characteristics along with their definitions.

\subsection{Instagram Posts on Journaling Spreads}

\textbf{Data Collection}. Instagram posts of journaling spreads were identified using popular hashtags such as ``\#journaling,'' ``\#journalinginspo'', and ``\#journalingspreads.'' We selected posts by prioritizing for a diversity of visual styles and the variety of materials incorporated in the journaling spreads. We collected 50 Instagram posts by 44 unique accounts (displayed in Table \ref{tab:instagram-posts}). Posts could include one or more images of journaling spreads. We excluded Reels and videos to avoid overlap with the YouTube videos. Instagram accounts from our dataset had a mean follower count of 39.8k (SD = 98.7k). Pictures and captions from these posts were downloaded and transferred to a shared database for analysis.

 \begin{figure*}[htb]
    \begin{center}
    \includegraphics[width=\textwidth]{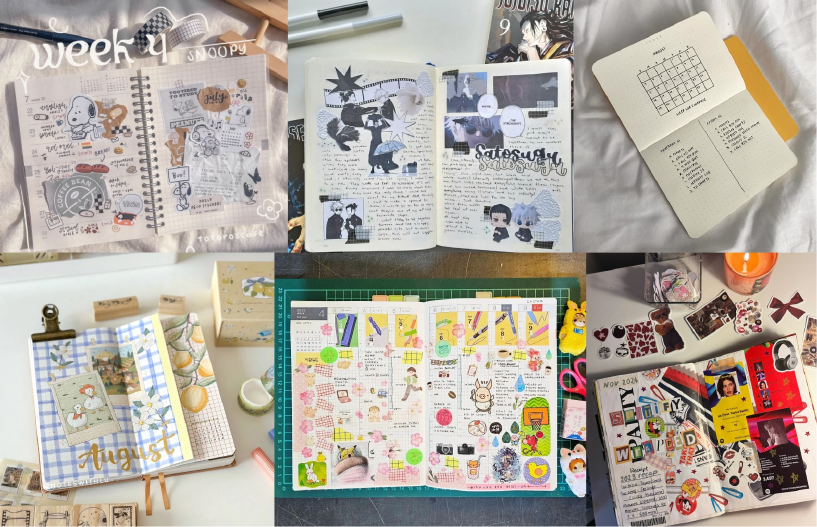}
        \captionof{figure}{Images from Instagram journaling spread posts analyzed for layouts (clockwise from top left: $@$totoroscafe, $@$k4rasu.\_, $@$ellensjournals, $@$isuris\_diary, $@$5am.raining, $@$noteswithlu)}
        \Description{Six images of Instagram posts with journaling spreads} %
        \label{fig:instagram-spreads}
    \end{center}
\end{figure*}

\textbf{Analysis}. Three researchers initially analyzed the same 10 Instagram posts by independently annotating images from each post, paying attention to the layout of the spread and their decorative aspects. After discussing and agreeing upon common attributes based on our collective annotations, the researchers added these open codes to a new affinity diagram. The remaining posts were then split up between two researchers, with each researcher annotating 20 additional posts. These annotations were then utilized to add onto existing codes, create new codes, and reformat them accordingly. Some of the codes concerned elements placed directly onto the journaling spread, such as stickers, washi tape, or cutouts intended purely for aesthetics, whereas others captured page layout. Our interpretations were discussed in weekly meetings over 2 months, at which point we added codes pertaining to spread layouts to the existing codebook. 

\subsection{Interviews on Journaling Practice}
\textbf{Recruitment.} To better understand user motivations for journaling and their journaling routines, we recruited 11 journalers to participate in semi-structured interviews. Participants were all students recruited from a large private research university in New York City. We advertised our interviews across campus using a combination of flyers and forwarded email invitations to groups and individuals who journal. We selected participants to balance across types of journaling practice (such as travel journaling, scrapbooking, reflective journaling), use of stationery accessories, and digital journaling experience. Our research protocol was reviewed and approved by our university's Institutional Review Board. 

\textbf{Protocol.} Participants selected for our interviews were invited to participate individually in a 60 minute in-person session with our research team. They were asked to bring a journal they would feel comfortable sharing and any of their favorite stationery, writing tools, or accessories. In our research, we defined a journal as a collection of entries that capture records of one’s experiences, thoughts, aspirations, and ideas. During our interviews, one researcher facilitated the session, while another observed and took notes.

The interviews utilized a semi-structured protocol incorporating questions about participants' general journaling practices (their motivations, their current journaling routine, and how their practices have changed over time) along with a walkthrough of at least one spread from their journal. The questions asked during guided walkthroughs were informed by the existing code book developed from our social media analysis; for example, we asked participants to elaborate on the motivations for their spreads, the use of certain tools, and their layouts. We also asked whether they share their journaling practices or spreads with others, engage with stationery brands, or are involved in journaling communities. We then inquired about their experience with digital tools, including using them in tandem with analog tools or as a primary method for journaling (for those who choose to not use digital tools, we asked them to compare the affordances of analog and digital journaling). Finally, we asked whether they had ideas for what could improve or support their journaling practice. Each participant received a \$20 gift certificate. 

\textbf{Analysis.} After each interview, the researcher who observed the interview wrote a memo on key points and themes of the discussion, including relevant quotes and photographs of participants' journaling spreads. Audio recordings from the interviews were transcribed, reviewed, and formatted for analysis. The transcripts were divided among four researchers, with each researcher performing an open coding pass for salient ideas relating to journaling practices and motivations. Open coding was used rather than applying existing codes from our code book on publicly shared journaling content as our interviews touched upon themes not addressed in publicly shared content, especially around how an individual's journaling practice developed and comparisons of analog and digital tools. Through weekly group research meetings, we discussed our interpretations and used affinity diagramming to identify and organize a new set of themes around journaling practices.

\subsection{Consolidation of Findings} %
After we had developed affinity diagrams for each data source (YouTube, Instagram, and interviews), we consolidated all identified themes into a single code book. We then performed a second pass on the interview transcripts, with each transcript reviewed by a researcher who did not conduct the first open-coding pass. This second pass involved the application of the consolidated code book, including identification of common themes across all data sources as well as the addition of new themes. We added participant quotes to our affinity diagram for our interview data, using this to ground group discussions and revise our final code book so that it incorporated themes from all data sources.

\subsection{Analog Interview Participants}
Here we provide a summary of our interview participants to give context to our overall findings in the next section. We interviewed 11 journalers (9 female and 2 male) between the ages of 18-25 (mean = 21 years old) who are all students at a large private research university in New York City. Our participants each had between 1-5+ years of journaling experience. Table \ref{tab:journalers} summarizes our participants (names given to journalers are pseudonyms) including their favorite stationery item along with any digital tools they had used for journaling. All participants consider analog journaling to be their dominant journaling practice, even if they had experience with digital tools -- the only exception was Ryan, who had previously journaled on paper but had switched to journaling with Obsidian, a digital note-taking application. 

\begin{table*}[t]
  \centering
  \caption{Analog Journalers Who Participated in Our Interviews}
  \label{tab:journalers}
  \begin{tabular}{llll}
    \toprule
    Participant & Years of Experience & Favorite Stationery Item & Digital Journaling Experience \\
    \midrule
    
    Lydia & 3-4 years & N/A & Notion, Apple Notes, Goodnotes \\
    Tara & 5+ years & Tombow Brush Pens & N/A \\
    Yasmine & 5+ years & Washi Tape & Goodnotes \\
    Abby & 1-2 years & Pilot G2 Pens & DayOne, Apple Notes \\
    Sophie & 5+ years & Sharpie Pens & Apple Notes \\
    Ava & 4-5 years & Cards and Fun Stamps & Apple Voice Memos, Apple Notes \\
    Aaron & 2-3 years & Origami Paper & Apple Notes, Procreate \\
    Ryan & 4-5 years & Zebra Multicolor pen & Obsidian, Apple Notes \\
    Beth & 5+ years & Sailor's Fountain Pens & Goodnotes \\
    Sadie & 5+ years & Stickers & Google Docs \\
    Alaina & 4-5 years & Stickers and Post Cards & Apple Notes \\
  \bottomrule
\end{tabular}
\end{table*} %
\section{Results}
\label{sec:results}

In this section, we begin with a summary of journaling characteristics we observed in our data, including an overview of analog stationery and digital tools used in journaling, along with an overall description of practices represented by participants in our interviews. We then describe how an individual's decisions about how, why, and what to journal (and how these decisions change over time) stem from the interplay among three factors: material choices, personal context, and communities, which we collectively refer to as the \textit{journaling ecosystem}. For each of these factors, we provide examples of how they manifest in journaling decisions, with particular note of how they influence changes in practice over time.

\textbf{Analog Stationery.} Journalers use a wide variety of stationery, including types of paper, notebooks, writing instruments (e.g., fountain pens, markers, gel pens), and accessories (e.g., stickers, washi tape, stamps). In addition to purchased products, accessories also include items like ephemera, or found paper-based objects that journalers can incorporate onto the page as memorabilia (like train or show tickets) along with photographs a journaler might print themselves. The combination of these materials are used for personalization and customization, expressing one's aesthetic.

\textbf{Digital Tools.} Through our Instagram analysis, we observed how portable digital photo printers are popular accessories for journalers to print photos for their spreads. Since our interviews allowed for conversations about how journalers considered digital tools, they surfaced the use of 1) supplementary digital tools (e.g., Google Calendar, Photos) to review and reflect on events to journal on paper about; 2) digital notetaking apps used in conjunction with or in support of physical journaling (e.g., Voice Memo, Apple Notes); and 3) digital apps meant to be primary journaling tools (e.g., One Note, Notion) that participants tried out. All journalers use a variety of digital tools, even if for most, their ultimate documentation lives on paper.

\begin{figure*}[htb]
    \begin{center}
        \includegraphics[width=\textwidth]{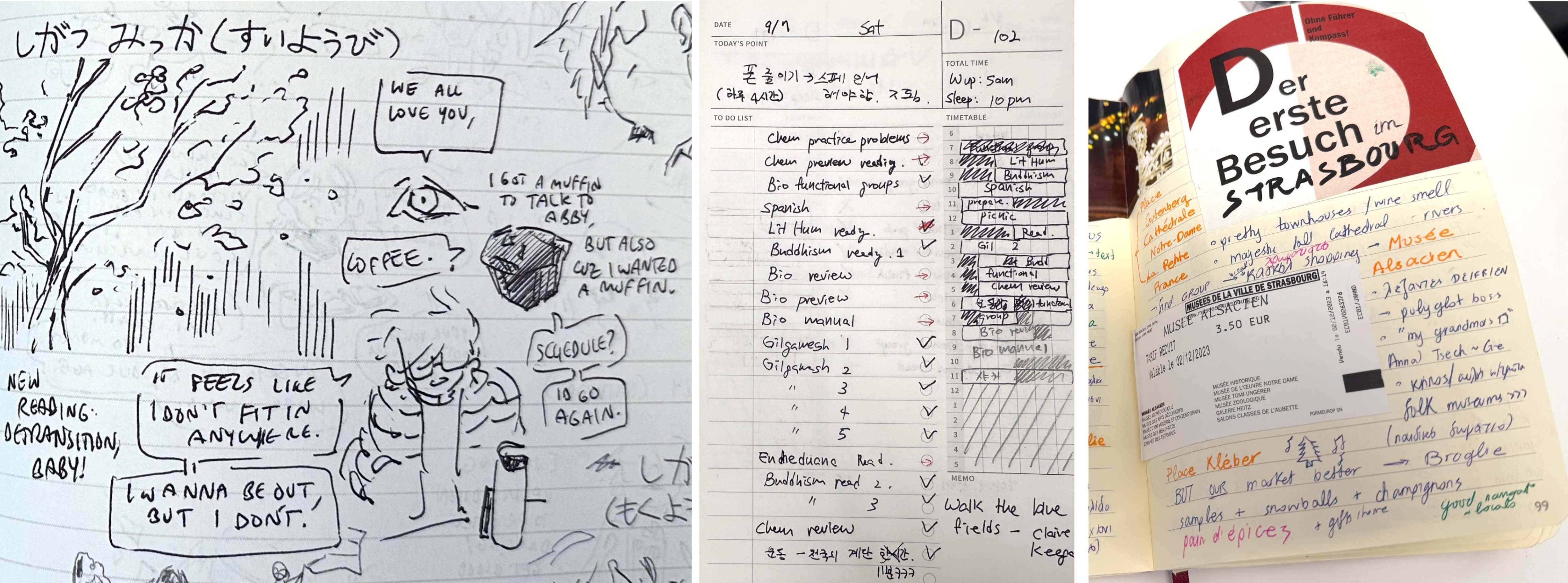}
        \captionof{figure}{Examples from our interview participants' physical journals (from left to right: Aaron, Beth, Alaina)}
        \Description{Three spreads that showcase art, templates, and ephemera based journaling} %
        \label{fig:participant-spreads}
    \end{center}
\end{figure*}

\textbf{Overview of Interviewed Journalers' Practices}. Our interview participants shared  spreads from both physical and digital journals, a selection of which are displayed in Figure \ref{fig:participant-spreads}. Interviewed journalers engage in a range of types of journaling, including art or visual journaling, bullet journaling, reflective journaling, scrapbooking, and travel journaling.

\subsection{The Journaling Ecosystem}
\begin{figure*}
    \centering
    \includegraphics[width=\textwidth]{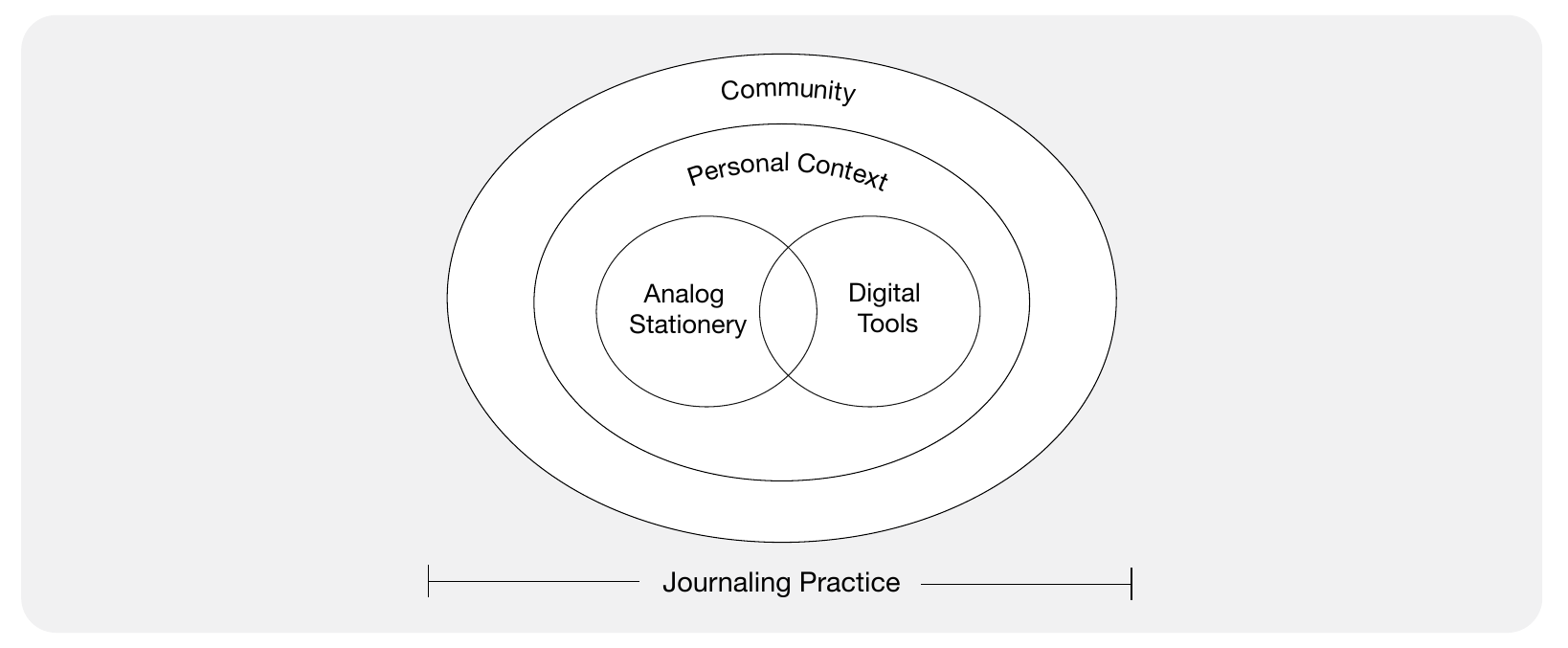}
    \caption{Journaling Ecosystem consisting of materials, personal context, and community.}
    \Description{A diagram showing intersection analog stationery and digital tools, surrounded by personal context, which is all surrounded by community} %
    \label{fig:ecosystem}
\end{figure*}

Our analysis revealed how personal choices for journaling practices are shaped by an ecosystem consisting of \textit{Materials} (both analog and digital), \textit{Communities}, and \textit{Personal Context}, as displayed in Figure \ref{fig:ecosystem}:

\begin{itemize}
    \item \textbf{Materials}: The \textit{stationery} (paper, journals, writing instruments, and accessories) that make up what journalers put down on the page, along with \textit{digital tools} that support or shape analog journaling practices. These tools can be used exclusively, or in tandem with one another.
  \item \textbf{Personal Context}: Personal circumstances in a journaler's life that influence the content they journal about and their journaling routines. Personal context informs both material choices and journaling practices.
  \item \textbf{Community}: People and communities with which an individual interacts with through journaling and gets inspiration or reference examples from. This aspect of the ecosystem includes both those in an individual's immediate environment as well as broader journaling communities (such as on social media).
\end{itemize}

In the following sections, we characterize each component of the journaling ecosystem, noting how they influence how people's practices develop over time as their priorities shift. Note that while we had initially anticipated our research focusing on user preferences for journaling tools, we found the stories of changing practices and preferences over time to be the most compelling, which meant the bulk of our results stemmed from our interviews rather than our initial social media analysis. However, our social media analysis provided an analytical lens that we applied to our interview protocol and enabled us to compare the aspirational journaling shared online with the more regular practices of our interviewees, specifically around journaling tool selection and layout design. (Throughout the Results, when referring to social media content, we include a footnote for attribution; otherwise, quotes are from participants in our journaling interviews.)

\subsubsection{Materials}
Material decisions around stationery and digital tool selection impacted ergonomics that help sustain and motivate a journaling routine.

\textbf{Reducing friction between thoughts and the page}. For journalers, the ability to seamlessly transfer their thoughts to paper motivates an appreciation for products that reduced friction both physically and cognitively: ``What is the smoothest writing notebook and smoothest writing pen that I can bring together for the ultimate journal experience? I wanted something that had as little friction as possible between my thoughts and what I was putting down on the page'' [Antonia from $@$JetPens]\footnote{\url{https://youtu.be/9DRm7m8vpd0?si=05RmFXGZXBJ_z091}}. Journalers described the ways certain combinations of materials can minimize distractions, including the use of dotted grid paper (``It's not very aggressive and so it gives you a little more freedom'' [Tara]) or neutral colored pens (``[brown ink is] softer on the eyes and I prefer it just for my everyday writing'' [$@$KaitlinGrey]\footnote{\url{https://youtu.be/tlNrCMbGWUM})}.

\begin{figure*}[htb]
    \centering
    \includegraphics[width=\textwidth]{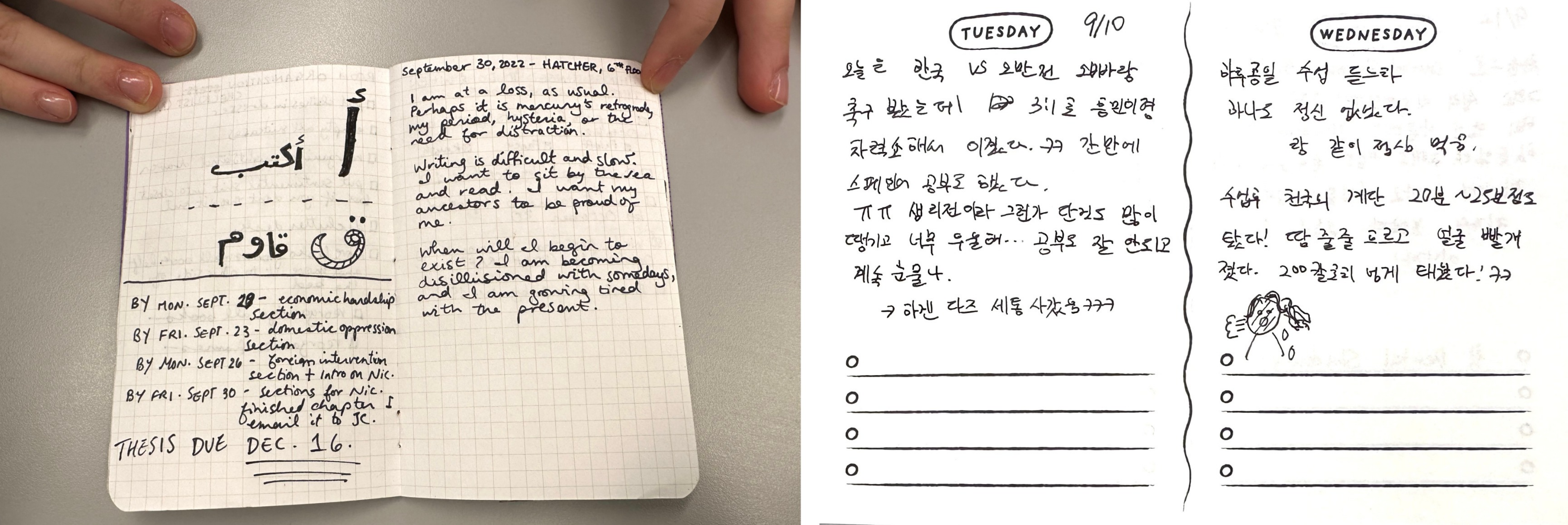}
    \caption{Example spreads of space constraints within physical journals (left: Sophie, right: Beth)}
    \Description{Two spreads that highlight space constraints within a journal. One spread has a grid paper and is more horizontally constrained and another spread with blank paper and a template that is more vertically constrained.} %
    \label{fig:space-constraints}
\end{figure*}

\textbf{Size shapes thought}. Several journalers shared how physical constraints, specifically on page and notebook size, helped minimize anxiety about getting started, while also reducing opportunities to overwrite. Multiple participants shared that an abundance of blank space can overwhelm and create pressure to fill the page. Beth stated how choosing a notebook with smaller space to write prevents her from emotional venting: ``I get overwhelmed by writing too much for myself.'' Similarly, our discussion with Sophie revealed how a pocket-sized notebook reduced fear: ``Having a small everyday journal helped me get over the fear of filling in a new notebook.'' In contrast, some journalers described how digital notetaking apps often present a blank canvas they feel obligated to fill. Journaler Abby shared how in the Day One journaling app, ``the space was just a big blank thing [and I] would turn [it] into a list of things I was upset about, which would make me feel worse.'' In comparison, she limits herself to a sentence or two in her paper journals.

\textbf{Enabling time to process}. Journalers appreciated slowness and deliberateness in writing, which can help them mentally process their thoughts: ``The physical act of writing down something makes you actually process things slower because you're not just typing whatever comes to mind -- you have to actually reflect on what you physically can write...given that it's much faster to type [than to write]'' [Tara]. 

\textbf{The role of editability.} While the ability to undo is common in digital tools, a side effect of this feature is that it can encourage perfectionism rather than more freeform writing. For example, Aaron, whose paper journal primarily consists of hand-drawn sketches, shared how analog drawing with ink has a permanency compared to drawing in the app Procreate: ``I might put more effort [on Procreate] into composition because you can undo easily. Whereas with the physical journal, it's always in pen, and it's always like, `Oh, [I] drew that. That's the end.''' Sadie would initially type out her journal entries, but discussed how a sense of perfectionism would impact her journaling: "I guess I was just trying to make it more perfect because I was typing so you can delete stuff." Critically, this instinct to perfect through editing may stand in opposition to stream-of-conscious writing that many journalers engage in.

\begin{figure*}[htb]
    \begin{center}
        \includegraphics[width=\textwidth]{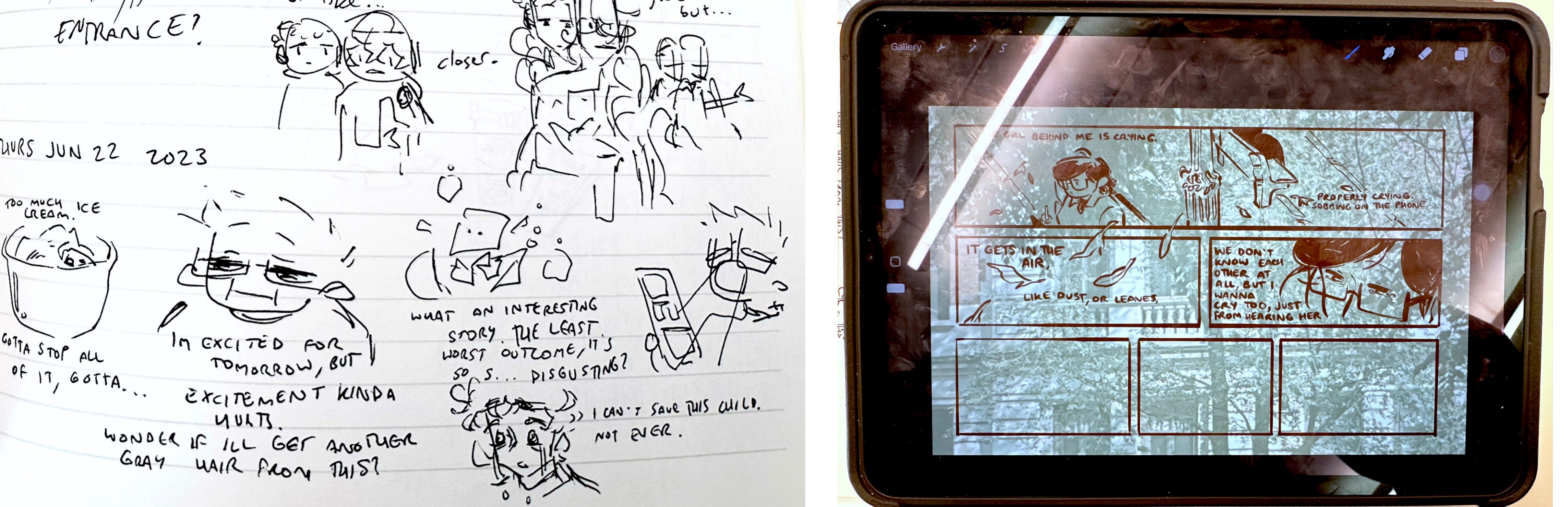}
        \captionof{figure}{Comparison of analog notebook style from Aaron (left), where he does not edit, and iPad Procreate drawings with high composition and heavy editing (right)}
        \Description{Two spreads that highlight editability and cleanliness of digital journaling versus analog journaling. One spread is art journaling done by hand in a notebook while the other is done digitally on an iPad.} %
        \label{fig:editability}
    \end{center}
\end{figure*}

\textbf{Aesthetics bring joy}. Aesthetics contribute to a journaler's decision to use particular stationery since it enables customization and motivates journaling practices. As seen in Figure \ref{fig:instagram-spreads}, our Instagram analysis identified that visual aesthetics are a major aspect of journaling practices and spread design. Our YouTube and interview analysis further revealed that analog tools, such as pens, notebooks, and accessories, that are visually appealing will uplift a journaling experience and motivate users to continue writing: ``signing my name has never been cuter [...] the expressions [of a character on the pen] change when you press the [clicker] — How cool is that?'' [$@$miicchaa]\footnote{https://www.youtube.com/watch?v=2Z3B-v2cugo}. Beth stated how she gets joy from journaling with and in pretty materials: ``A really pretty pen or writing [in] a pretty journal like planner makes you just happy by itself just to look at it.'' For Sadie, accessories uplift and serve as a reward for her hard work or writing: ``It's [using stickers] just to make myself feel better. It's so cute -- it just makes me happy.''

\textbf{Combining tools in different contexts}. Journaling spreads on Instagram revealed that journalers use a combination of digital tools, (such as Spotify and Photos as seen in the bottom right image of Figure \ref{fig:instagram-spreads}), in tandem with analog tools, such as ephemera and stickers, to create their spreads. Journalers in the interviews also described how note-taking apps like Apple Notes and Voice Memos can act as ``staging'' grounds for taking notes that later get translated into physical journal entries. Alaina and Ava, two journalers we interviewed, described how these applications served as a place to jot down intermediary thoughts when their physical journals are not accessible, like on a train. In this way, users could take advantage of mobile devices to quickly capture their thoughts, which they would then manually review to determine what was worth capturing in their journals at a later date. This practice suggests the value of convenience in supporting regular, intermediary notetaking for content that eventually ends up on the page.

\textbf{Material choices over time: Digital tools reconfiguring journaling routines}. For some interview participants, digital tools eliminated certain journaling practices that they previously engaged in. For Ryan and Alaina, the addition of Google Calendar into their routine ultimately replaced the ``planning'' oriented writing that their journals used to have, such as for event and task tracking. This, in turn, altered their journaling to primarily consist of reflective entries about the past.

\textbf{Material choices over time: Crafting one's own style}. Page templates are layouts built into notebooks that often have labeled sections such as checklists for tasks, writing prompts, and grids for calendar entries. These templates can offer guidelines to adhere to when writing, but several journalers described how they eventually moved beyond pre-made templates. For example, Abby originally used a notebook with specific writing prompts (like the best food she ate that day), but over time, she felt that the prompts became stale, making journaling feel like a chore. She transitioned to using a blank notebook in which she created her own categories and now adds and removes categories depending on how well they work for her. For example, she has monthly `gratitude' and `reflections' sections and experimented for half a year with an `intentions' category for setting goals; however, she eventually removed `intentions' because she felt a lack of control over the outcomes of many of the items she tried to plan for, leaving her feeling annoyed and negatively impacting her outlook. Similarly, Lydia, who uses Notion for digital journaling, initially started with a complex bullet-journal-like template with multiple subsections but eventually opted for a simpler, largely blank one because the original felt too structured. Through developing and refining their own templates, these journalers are continually crafting their own practices based on what they find most conducive to a positive routine.

\subsubsection{Personal Context}

A journaler's motivations and practices are often shaped by where they are in their lives and who they hope to become. On a practical level, journaling routines are also influenced by one's life situations, including access and competing priorities.

\textbf{Development of identity}. Our interviews showed how a journaler’s identity is reflected in their practice, with journaling supporting reflection on one's personal journey and development and serving as a means to improve valuable skills related to their identities. Participants enjoyed looking back at old entries to reflect on past versions of themselves, see how their values have changed over time, and apply past insights to current situations. Since many of our interviewees had been journaling since their youth, participants like Sophie shared how she found humor in revisiting what was upsetting to her when she was a teenager. Lydia expressed a desire for a journal that could help her revisit past entries, with the hope of addressing current dilemmas: ``I want to see [how] I reacted in the past because I feel like people always make the same mistake.'' The ability to revisit her past through her journal could thus serve as a means to reflect and potentially adapt her current behavior based on past outcomes. The tools journalers use to support reflection in their practices also demonstrate the value and desire to revisit previous entries: ``I feel like it [photos from a Instax Mini Link Printer] [...] adds an extra element of nostalgia when looking back at my entries ''[$@$KaitlinGrey]\footnote{\url{https://www.youtube.com/watch?v=yUebKmgQVvM}}.

Examples of reinforcing identities include Sophie who, as a previous history major, values her writing and ability to communicate ideas for the purpose of historical preservation. She noted how journaling has helped her academic writing through being a medium ``to practice [writing], just like you would a sport or an instrument.'' For art major Aaron, art journaling has helped him develop his sketching skills: ``I took some of this and used it within art I was doing on Twitter...it's also [helpful for] sort of developing compositionally and practicing.'' For these journalers, their practices directly reflect and impact key aspects of their identity by supporting development of highly valued skill sets. 

\textbf{Shaping mindset}. For many, journaling was seen as a tool to promote positive outlooks on life and improve mental health. Participant Ava described her journal as being a ``sidekick'' to help her get through the day, representing a space for providing encouragement to herself and reminding her of the good in life. Shifting her practice from emotional venting to gratitude journaling, Ava shared how this change was motivated by a realization that her previous journaling practice was worsening her mental health: ``When you write [problems] and see them, [you] kind of believe them...complaining about things, it almost amplifies the problem that you're facing.'' This sentiment was echoed by Beth: ``When you journal and write it down, you're doubling the memory, so it's gonna stress you out twice.'' Alaina, who described a natural tendency to focus ``on the bad,'' similarly spoke about how she used journaling to ``become a less negative person,'' sharing, ``It's [her journal's] like a therapist... It has a very big impact on my mental health.'' Rather than using her journal to expound on a particular problem, she has a preference to document concrete steps for working towards addressing or resolving issues, even if this may not fully represent challenging moments in her life, largely because it helps shape a positive outlook and motivates her to keep progressing. 

\textbf{Personal contexts over time: Reconfiguring journaling routines}.
Going to college typically results in major shifts such as becoming more independent and sometimes moving countries and acclimating to new cultures. Many of our participants' journaling practices changed from pre-college in response to these life changes. They described how while they previously had binders of stickers and other accessories at home, moving into a smaller, shared space (like a dorm room) meant they had to be more selective about the accessories they used due to space constraints. For others, finally having their own space meant more freedom because of reduced risk of family members reading their journals.  A shifting emphasis on functionality in journaling among participants was most likely related to the busy schedules students need to maintain and an overarching need for efficiency. Their identities as college students also factored into their economic choices of what products to use, with participants like Sophie switching to more affordable Muji notebooks, and Beth sharing the potential of digital journaling tools as a way to alleviate the financial burden of expanding her sticker stationery collection.

\subsubsection{Community}
While journaling preferences can be highly personal, journaling itself is not an isolated process. People's relationships with others can inform how they journal, and social expectations (such as through journaling communities on social media) can impact practices people try for themselves.

\textbf{Developing relationships through journaling}. Interviewees spoke to how the act of journaling together can foster stronger relationships and create moments of shared connection. Ava described how a communal journaling session with friends created a way to learn about others' preferences, describing how her friends shared prompts they used in their notebooks and how ``it was so cool to see how we all navigated it [journaling] differently.'' Sophie leverages her journal as a tool to develop relationships through incorporating her friends directly into the process, both through co-drawing diagrams in the same notebook and recording and sharing memorable quotes from friends. Ava frequently takes pictures of her notebook and shares these entries with loved ones, including quotes or memories that express her appreciation for them. 

In addition to building relationships through lighthearted moments, we also discussed situations where journals were a means to resolve inter-personal conflicts. Ryan uses his reflective writing to gain third-person opinions on conflicts he has with others, using his entries to provide context and capture his perspective without directly sharing the journal itself. The ability to use the journal to organize his thoughts and address inter-personal conflicts points to journaling as a powerful tool for continuous relationship development.

\begin{figure*}[htb]
    \begin{center}
        \includegraphics[width=\textwidth]{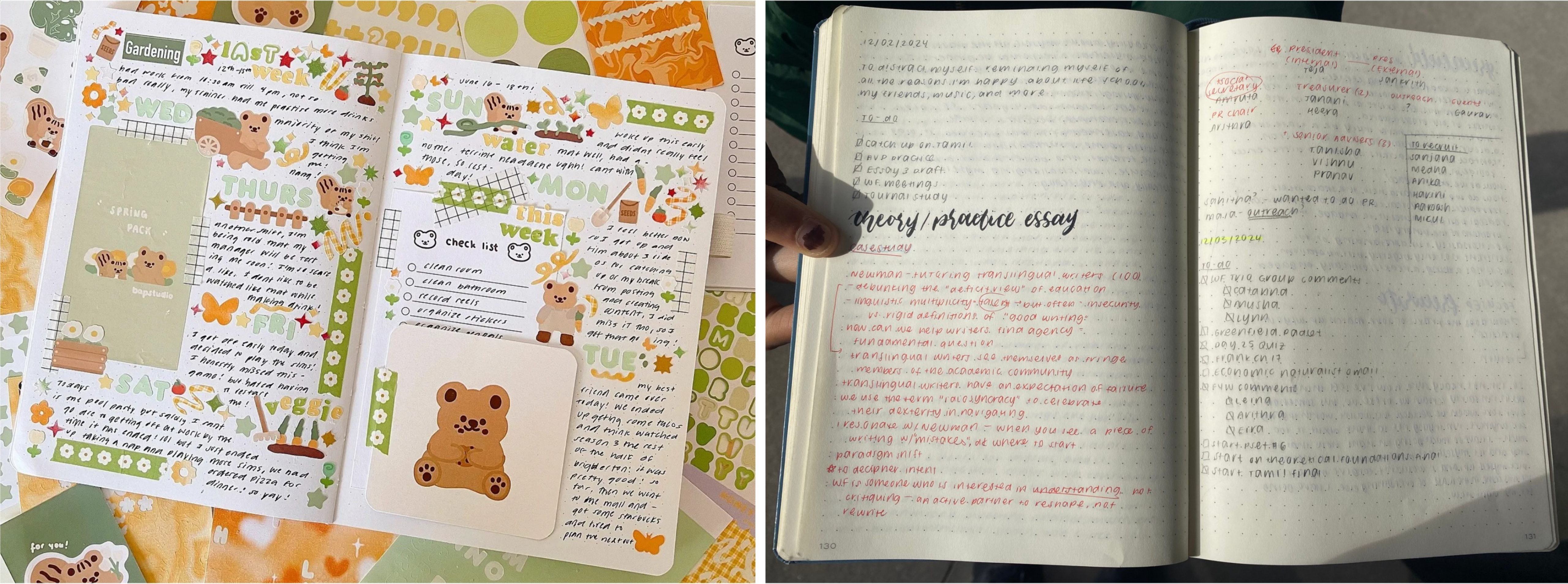}
        \captionof{figure}{A highly-decorated journaling spread shared on Instagram by $@$shan.jrnls and $@$bapxstudio (left), compared to a spread from Tara (right) who shared her reservations about aesthetic journaling after trying out the practice for herself}
        \Description{Two spreads that demonstrate aesthetic journaling in the public data versus interview data. One spread is an example from Instagram that uses lots of stickers and colors while the other spread is done in only pen and pencil with very minimal use of colored ink.} %
        \label{fig:aesthetic-journaling}
    \end{center}
\end{figure*}

\textbf{Community over time: Reappraising social expectations}. We observed in our Instagram data how many journalers who post online can take pride through artful construction of meticulously decorated spreads that incorporate sketches, calligraphy, stickers, and collages (as displayed in Figure \ref{fig:aesthetic-journaling}). Our interviewees shared a different reality--while a number described seeing examples of aesthetic journaling on social media and shared how these exemplars motivated them to try out this approach themselves, they largely moved away from these methods, ultimately seeing it as not authentic or sustainable. Ava detailed how this style of journaling was aspirational but ultimately incompatible: ``[I thought] `Oh, I can be that girl that has a really nice journal and her thoughts are so collected and it's very beautiful and pleasing’...the reality of it was that I didn't. Now I think I'm much more more authentic.'' Authenticity was also touched upon by Sadie, who shared, "Everything in this journal is a little haphazard, but that's also how all my writing is, so you gotta keep it authentic. Like, I'm not gonna lay out this beautiful spread. I also just don't have time for that.. it's not going to work for me.'' In the end, accessorizing did not feel true to the ideas they were trying to express: ``I feel like if I was to like add stickers and tape and whatnot, for me personally it feels a bit forced and that like I'm worried more about the aesthetics versus the actual substance'' (Abby); ``I felt like decorating my journal...it's not really expressing myself anymore. For me, it's more about putting the words down instead of `oh there's a flower, there's an animal [sticker]'" (Yasmine). These examples show how while social media can depict high expectations for journaling style and time commitment, journalers we spoke to claimed their own priorities based on what felt authentic to themselves.

\section{Discussion}
\label{sec:discussion}

Our analysis revealed the ways in which journaling practices, including preferences for tools and desires for sustainable routines, are continually formed through an ecosystem beyond the page. Here, we comment on potential design opportunities that might incorporate and accommodate these changes to support the personal journey of crafting one's journaling practice.

\subsection{Accommodating Changing Needs and Preferences}
Our interviews revealed how journalers adapt their practice in response to changes in personal context, priorities, and needs. This process involves identifying what types of content are meaningful and productive for themselves over time. For example, Abby and Lydia both began with structured, pre-made templates, but found fixed prompts and numerous sections to be overly prescriptive. Instead, they came to craft their own templates with sections that better matched their goals. Similarly, while some of our interviewees were encouraged to try out ``aesthetic journaling'' they encountered on social media, they quickly found that the heavy use of accessories felt inauthentic to their own voice. Yet, even if existing templates were ultimately abandoned, they played an important role in shaping personal preferences through the identification of practices that do not work well for one's goals.

We also observed how, rather than using journaling as a way to ``account for one's life'' and objectively capture what an individual did as found in previous work \cite{elsden2016s}, journalers often selectively captured more positive moments in their life over negative ones. For these users, journaling was less about serving as an objective measure and more about shaping a positive outlook and mindset. As Beth described, writing down stressful events on the page ended up reinforcing negative thoughts: ``You're kind of like doubling the memory, so it's going to stress you out twice.'' Instead, journalers like Alaina and Ava elected to focus on positivity and gratitude journaling, even though they acknowledged that it did not necessarily capture the full emotional spectrum of their experiences. The filtering of content thus became a method of self-preservation, decreasing anxiety about reflecting on one's past and reducing hesitation to revisit one's journaling practice.

These results suggest that the design of reflective journaling tools should adapt and support changing practices over time, while also not necessarily prioritizing strictly accurate documentation of one's life. Crafting one's journaling preference is an important journey towards choosing and representing dimensions deemed worth reflecting on and that will positively support sustained journaling practice. Notably, the journalers we spoke to that used digital tools predominantly preferred more general-purpose notetaking applications (Apple Notes, Notion, Obsidian) as opposed to journaling-specific apps (only one shared that she tried Day One before abandoning it); these note-taking apps may offer more flexibility as they do not prescribe a particular template and thus may allow for more customization.

\subsection{Helpful Constraints for Cultivating Mindful Journaling Practices}
Journalers shared how material and physical constraints helped support their practice. Small notebooks and space-constrained pages made it less intimidating to write, reducing journalers' fear of starting a new notebook or facing a blank page. Contrasting these limits with digital applications, journalers described how having an unbounded text box in applications like Day One was underconstrained, leading them to write too much, especially outpourings of negative thoughts, which could increase stress. Limited space on the page helped users create natural boundaries for containing their thoughts, which meant both less venting and reducing the amount of time for their journaling routine, encouraging a more sustainable practice.

Further, while the ability to undo in digital apps can be seen as beneficial from an editing perspective, journalers can find this freedom overwhelming when it encourages perfectionism. With stream-of-conscious journaling, users focus more on putting ink to the page and seeing where their ideas go, as opposed to crafting a perfect version of their thoughts. When comparing the digital drawing app Procreate with physical notebook sketching, Aaron described how the ease of undo in Procreate led him to spend much more time perfecting his spreads, whereas with physical journal, he chooses not to erase or cross out any of his work, supporting his flow. In this way, the permanency of ink may encourage unfiltered writing.

These helpful constraints suggest that in contrast to features in digital notetaking apps today, providing bounded space for writing, along with encouraging more freeform, unedited thoughts through removing or deemphasizing undoing, can be valuable characteristics; however, further work is needed to determine the right balance that still accounts for digital affordances that users may value, such as with Gulotta et al's work on user perceptions of incorporating physical decay into digital photographs to mixed user reactions \cite{gulotta2013digital}.

\subsection{Navigating the Larger Journaling Ecosystem}
Evolving journaling practices described by our interviewees revealed how changes stem not just from the individual themselves, but their relationships with a larger ecosystem of external factors.

While \textit{digital tools} like calendar applications are not used for journaling per se, their role in keeping track of daily events led journalers to reduce or remove planning-oriented activities from their analog journaling routine, focusing instead on reflective journaling.  Further, digital tools served as a means to \textit{remember}, with reviewing digital calendars and photos being a common practice in advance of writing on paper.  Given that documentation of modern life tend to sprawl across multiple digital platforms, there are opportunities to consider how these streams might be thoughtfully integrated to align with reflective journaling practice. As curation of one's life is a critical step in determining what to journal about, how might future tools help surface important moments to write about? As one example, the Apple Journal app \footnote{https://apps.apple.com/us/app/journal/id6447391597} automatically utilizes user activity on their mobile device to suggest topics for journaling entries, such as podcasts listened to and photos taken. However, given user preferences for digital detoxing and valuing the use of analog tools, how might analog and digital tools work together to extend rather than replace analog practice?

Second, we found that \textit{communities}, specifically journalers' relationships with external communities, differed from more recent work, including research on quantified-self journaling \cite{abtahi2020understanding, choe2014understanding}, where communities may more commonly have shared goals with trackable data, and blogging\cite{davis2014homemade}, where an external audience is implied. What our interviewees chose to write about was often highly personal and not commonly shared as is, at least not on social media. Instead, those who shared their journals did so with close friends who could encourage their practice, or as a place to organize thoughts to work through interpersonal conflict. Sharing with a small group of confidants, rather than public sharing observed with other journaling communities, may suggest that journaling tools should consider unique needs for helping users curate what is shareable, and how this may be done in with select individuals.

Third, a journaler's \textit{personal context} affected attributes like how frequently they journaled, what stationery they chose to use, and what they wrote about. In relation to changes over time, they described how their journaling routine differed from pre-college because of space constraints and less time to journal, leading them to opt for shorter journal entries and irregular writing routines. These changes in response to personal context led us to consider how journaling tools could adapt to one's current state, such as being aware of when a user is especially busy, or has a considerable life change that might need different types of support--ideas we hope to explore in future work.

\section{Limitations}
\label{sec:limitations}

Our journaling interview participants were all university students, and while they represented a range of journaling practices, it is possible that students may differ in several ways from journalers more broadly, especially those we observed on social media. For example, students may have more limited expendable income to spend on stationery, or be more likely to have less time to dedicate to crafting journal layouts. At the same time, we acknowledge that journalers sharing on social media may represent another end of the spectrum in which the time spent journaling is higher than average and more performative in nature. We believe that the mix of these perspectives from our analysis offer a valuable lens into a range of journaling practice, but interviews with an expanded set of participants (including non-students) could potentially shed additional light on user preferences and priorities.

\section{Conclusion}
\label{sec:conclusion}
Personal journaling is associated with many positive side effects, including promoting self-identity and well-being; however, in order to best sustain journaling practice, designers need to understand how personal journaling preferences and routines change over time depending on an individual's goals. In our work, we examined contemporary journaling through an analysis of social media posts about journaling and in-depth interviews with analog journaling enthusiasts, finding that journaling practice is shaped by an ecosystem consisting of material choices, personal contexts, and communities. By illustrating how each of these components of the ecosystem affect how and why people journal, we offer several design opportunities for future journaling tools, including better accommodating changing practices over time, supporting productive physical constraints, and adjusting sharing options to better support healthy relationships. We hope that these insights ultimately help tool designers in identifying opportunities to bridge analog and digital practices for enhanced personal documentation. 
\begin{acks}
We would like to thank Victoria Kirst for her valuable feedback about this work, along with all the journalers who participated in our interview study.
\end{acks}

\bibliographystyle{ACM-Reference-Format}
\bibliography{references}

\appendix
\clearpage
\onecolumn

\section{Appendix}

\begin{longtable}{lp{6cm}lp{4.5cm}}
  \caption{YouTube Videos in our Journaling Dataset}
  \label{tab:youtube-videos}

    \\ \toprule
    \# & Video Title & Creator & URL \\
    \midrule

\endfirsthead

\multicolumn{4}{c}%
{{\bfseries \tablename\ \thetable{} -- continued from previous page}} \\

\toprule
    \# & Video Title & Creator & URL \\
\midrule
\endhead

\hline \multicolumn{4}{r}{{Continued on next page}} \\ \hline
\endfoot

\hline \hline
\endlastfoot

    1 & What's in my Stationery Pouch? Delfonics Medium Pouch \& TSL Engineer Pouch
 & $@$JobsJournal & \url{https://youtu.be/-liZZQi4k14} \\
 
    2 & Journalling Pouch Tour All My Essential Journalling Supplies  & $@$KaitlinGrey  & \url{https://youtu.be/yUebKmgQVvM} \\

    3 & What's in my pen case? ft. Antonia's Everyday Carry   & $@$JetPens  & \url{https://youtu.be/9DRm7m8vpd0} \\

    4 & what's in my pencil cases stationery recommendations  & $@$amabelle  & \url{https://youtu.be/2_p2O9GTxbE} \\

    5 & what's in my stationery pouch my favourite planner \& journal supplies  & $@$hellysjournal  & \url{https://youtu.be/GetYI9Ezox0} \\

    6 & What's in my PEN CASE? ft. Stephanie + Corgi Sticky Notes \& MORE!  & $@$JetPens  & \url{https://youtu.be/xwNaK1jmGTY} \\

    7 & What's in Connie's Everyday Bag + Q\&A  & $@$JetPens  & \url{https://youtu.be/heg-sX11-nQ} \\

    8 & what's in my pencil cases  & $@$leighpadme  & \url{https://youtu.be/YZa4zGl7SzY} \\

    9 & what's in my pencil case 2024 uni essentials minimal stationary  & $@$elegant\_jen  & \url{https://youtu.be/6ruOjS4KOF4} \\

    10 & what's in my pencil case 2023  & $@$emmagrace.  & \url{https://youtu.be/EhHFDf0ZRik} \\

    11 & what's in my pencil case after 8 months of a no-buy  & $@$tbhstudying  & \url{https://youtu.be/Xlxs0mrkzsU} \\

    12 & what's in my pen case superior labor utility case my favourite journaling \& writing tools  & $@$KaitlinGrey  & \url{https://youtu.be/tlNrCMbGWUM} \\

    13 & What's In My Pencil Case?? (what stationery I bring when i'm travelling)  & $@$amandarachlee  & \url{https://youtu.be/8cPQkSbHyY4} \\

    14 & what's in my pencil case 2024 aesthetic stationery essentials for note taking  journaling  & $@$miicchaa  & \url{https://youtu.be/2Z3B-v2cugo} \\

     15 & What's in My Pencil Pouch 2024 Everyday Journalling Pouch, Travelling Case  & $@$MyLifeMits  & \url{https://youtu.be/A_i-h2p1vO8} \\

     16 & What's in my pencil case? My everyday + favorite stationery (for bullet journaling)\!  & $@$nouriyah  & \url{https://youtu.be/Ekjrd4joTD8} \\

      17 & What's In My Pencil Case Hobonichi Small Drawer Pouch  & $@$thegabypages  & \url{https://youtu.be/6r6Eaeb3FIg} \\

      18 & what's in my pencil case 2022 aesthetic stationery essentials for note takin journaling  & $@$cqndymlk  & \url{https://youtu.be/kdBA16FXhA8} \\

      19 & What is in My Pencil Case$?!$  Bullet Journal Tools I Love & $@$ErinFlotoDesigns  & \url{https://youtu.be/bs_R2PUmOTs} \\

      20 & what's in my pencil case 2024 aesthetic pinterest & $@$miinjjiii
  & \url{https://youtu.be/C_tbsNUwX4c} \\

      21 & What's in My Pencil Case (journaling, drawing and general writing) & $@$MariaTheMillennial
  & \url{https://youtu.be/ATQwvs5q3gA} \\

      22 & what's in my backpack as a schoolgirl: pencil case tour, muji binders, stationery, asmr, mildliners & $@$chloelizi
    & \url{https://youtu.be/dH7ElU_OGEM} \\

      23 & What's in my Journaling Pouch Delfonics Bag Tour & $@$sunnysideupjournals
    & \url{https://youtu.be/_SQeK0vWTUk} \\

      24 & What's in my Delfonics \& Mizutama multi stationery pouch (travel edition) & $@$onjerraslist
    & \url{https://youtu.be/aFvTpGHxH-c} \\

      25 & What's in my Journal Bag/Pouch? & $@$tisamuico
    & \url{https://youtu.be/MtodYhPwAj8} \\

      26 & POCKET EDC BULLET JOURNAL & $@$SigogglinJack
    & \url{https://youtu.be/0seLW2pHVek} \\

      27 & What's in my bag? The EDC of a Bullet Journal Guy & $@$MattRagland
    & \url{https://youtu.be/d685Ia1z4sk} \\

    28 & Field Notes Traveler's Notebook: What's in My Every Day Carry (EDC) and Travel Journaling & $@$retrowtures3896
    & \url{https://youtu.be/EEOmGria9aU} \\

    29 & what's in my pencil cases 2023 minimal stationery faves and recommendations & $@$vanillaa\_\_
    & \url{https://youtu.be/45AgJ7aawKE} \\

\end{longtable}

\begin{longtable}{lp{5cm}p{7cm}}
  \caption{Instagram Journaling Posts in our Dataset}
  \label{tab:instagram-posts}

    \\ \toprule
    \# & Creator & URL \\
    \midrule

\endfirsthead

\multicolumn{3}{c}%
{{\bfseries \tablename\ \thetable{} -- continued from previous page}} \\

\toprule
    \# & Creator & URL \\
\midrule
\endhead

\hline \multicolumn{3}{r}{{Continued on next page}} \\ \hline
\endfoot

\hline \hline
\endlastfoot

    1 & $@$paperplant.co & \url{https://www.instagram.com/p/C8fpxJnyZ7Q} \\
    
    2 & $@$5am.raining & \url{https://www.instagram.com/p/C0tVsutLFwd} \\

    3 & $@$diarylogs & \url{https://www.instagram.com/p/C06gTWeoIoz} \\

    4 & $@$amandarachlee & \url{https://www.instagram.com/p/C-WHHP9O5Fi} \\

    5 & $@$rishdrawings & \url{https://www.instagram.com/p/C7pBJDRJU29} \\

    6 & $@$noteswithlu & \url{https://www.instagram.com/p/C-JG5GEPijK} \\

    7 & $@$b.\_\_journalu & \url{https://www.instagram.com/p/C9U0_jkzF1z} \\

    8 & $@$every1luvsmika & \url{https://www.instagram.com/p/C-csmZnzEAB} \\

    9 & $@$bleuberry.ice & \url{https://www.instagram.com/p/C6joLprrLEy} \\

    10 & $@$offoffonoff2 & \url{https://www.instagram.com/p/C639edGP4Ci} \\

    11 & $@$shan.jrnls \& $@$bapxstudio & \url{https://www.instagram.com/p/C8ZztOAuuFF} \\

    12 & $@$artbrinana & \url{https://www.instagram.com/p/C9U8w3Fuons} \\

    13 & $@$\_marshmallowstudio\_ & \url{https://www.instagram.com/p/C_EiFKLyVYq} \\

    14 & $@$only.for.paperlovers & \url{https://www.instagram.com/p/C7Hz_UlIRdM} \\

    15 & $@$every1luvsmika & \url{https://www.instagram.com/p/C-NPzABTc9M} \\

    16 & $@$melodyglume & \url{https://www.instagram.com/p/C_h5ntdTuk3} \\

    17 & $@$kuroqnotes & \url{https://www.instagram.com/p/C_ib0GhIKo5} \\

    18 & $@$\_\_\_\_\_\_\_asa\_0905 & \url{https://www.instagram.com/p/C9XHkbRS0Us} \\

    19 & $@$stationeryproblems & \url{https://www.instagram.com/p/DAJaGfyRpTp} \\  

    20 & $@$6almy & \url{https://www.instagram.com/p/C_r8UGpPKN1} \\   

    21 & $@$highpaperclouds & \url{https://www.instagram.com/p/C3yHlZ6u5KI} \\   

    22 & $@$runaway.angels & \url{https://www.instagram.com/p/C9cesrryQrH} \\   

    23 & $@$mo\_mo\_ten\_log & \url{https://www.instagram.com/p/Cwpc2u0pDFb} \\ 

    24 & $@$k4rasu.\_ & \url{https://www.instagram.com/p/DAOrCazOz2v} \\ 

    25 & $@$abbeysy & \url{https://www.instagram.com/p/C3Z-qNHtKC6} \\ 

    26 & $@$salam\_creative\_ & \url{https://www.instagram.com/p/CZHEdmmoODT} \\ 

    27 & $@$totoroscafe & \url{https://www.instagram.com/p/C99nMU4TV-B} \\ 

    28 & $@$\_marshmallowstudio\_ & \url{https://www.instagram.com/p/C_oonNny1EL} \\ 

    29 & $@$sleepyneko.bujo & \url{https://www.instagram.com/p/DAVyBaavWG_} \\ 

    30 & $@$shinestickerstudio & \url{https://www.instagram.com/p/C4vqezaRGfo} \\ 

    31 & $@$euyos & \url{https://www.instagram.com/p/C9r5XgtzUQ4} \\ 

    32 & $@$mamagloriashop & \url{https://www.instagram.com/p/C-XwGHSIEU7} \\ 

    33 & $@$walkinginlavender & \url{https://www.instagram.com/p/C5tBSE3u34X} \\ 

    34 & $@$ara.byuuu & \url{https://www.instagram.com/p/C9_Wa3IPXA4} \\ 

    35 & $@$aurore.studio & \url{https://www.instagram.com/p/C5iukG3rLyD} \\ 

    36 & $@$ohnoelbujo & \url{https://www.instagram.com/p/C9AIPKRKrrV} \\ 

    37 & $@$teanbujo & \url{https://www.instagram.com/p/C6gWwwCK8ZA} \\ 

    38 & $@$my.artistic.journal & \url{https://www.instagram.com/p/C45ZD8dr_LZ} \\ 

    39 & $@$koook.ie & \url{https://www.instagram.com/p/C1uzD_GoUH3} \\ 

    40 & $@$\_theartof\_ & \url{https://www.instagram.com/p/C4-8teJKFBd} \\ 

    41 & $@$ellensjournals & \url{https://www.instagram.com/p/C-bTgv5R09-} \\ 

    42 & $@$rishdrawings & \url{https://www.instagram.com/p/C6cQWnASf5I} \\ 

    43 & $@$tsukiisamas\_shortcake & \url{https://www.instagram.com/p/C-U155JT9np} \\

    44 & $@$ceraiis & \url{https://www.instagram.com/p/C-dwAF_Ort-} \\    

    45 & $@$judi\_\_stone & \url{https://www.instagram.com/p/C8KPQ15qtcg} \\    

    46 & $@$sleepyneko.bujo & \url{https://www.instagram.com/p/C6JO6bXrNRP} \\ 

    47 & $@$journal\_\_junkies & \url{https://www.instagram.com/p/BYSOeMfgeCz} \\ 

    48 & $@$commonplace.log & \url{https://www.instagram.com/p/DAD6YMAoI2L} \\ 

    49 & $@$sharonajournals & \url{https://www.instagram.com/p/BmgSsIcn09J} \\ 

    50 & $@$isuris\_diary & \url{https://www.instagram.com/p/DDqkPLRIgYd} \\

\end{longtable}

\clearpage
\twocolumn 
\end{document}